\documentclass[twocolumn]{rspublic}
\usepackage{graphicx}
\usepackage{amssymb}
\usepackage{amsmath}
\usepackage{epstopdf}
\usepackage{subfig}
\newcommand{\half}{ \frac{1}{2} }

\newcommand{\OD}[2]{ \frac{d#1}{d#2} }

\newcommand{\PD}[2]{ \frac{\partial#1}{\partial#2} }

\newcommand{\PDO}[1]{ \frac{\partial}{\partial #1} }

\newcommand{ \Ket }[1]{ \left| #1 \right> }
\newcommand{ \Braket }[2]{ \left< #1 | #2 \right> }

\begin{document}
\title[GRAPE: principles and practicalities]{Implementing quantum logic gates with GRAPE: principles and practicalities}
\author[B. Rowland and J. A. Jones]{Benjamin Rowland and Jonathan A. Jones}
\affiliation{Oxford Centre for Quantum Computation, Clarendon Laboratory, Parks Road, Oxford OX1~3PU, UK}
\label{firstpage}
\maketitle
\begin{abstract}{NMR, quantum information, GRAPE pulses}
We briefly describe the use of GRAPE pulses to implement quantum logic gates in NMR quantum computers, and discuss a range of simple extensions to the core technique.  We then consider a range of difficulties which can arise in practical implementations of GRAPE sequences, reflecting non-idealities in the experimental systems used.
\end{abstract}

In developing a quantum computer, one of the most important requirements is the ability to perform any desired logic gate on the system (Bennett \& DiVincenzo 2000). In heteronuclear NMR implementations, where each qubit in a molecule can be addressed individually this problem is essentially solved. All single qubit gates may be performed trivially, using resonant radiofrequency (RF) pulses for rotations around axes in the $xy$-plane, and networks of such pulses for rotations around other axes (Jones 2001, 2011).  The two qubit controlled-phase gate required to complete a universal set of gates can be implemented merely be a period of interaction under the natural coupling term between a pair of spins. In systems with more than two qubits, unwanted interaction terms may be suppressed using spin echoes (Jones \& Knill 1999). Furthermore, composite pulse sequences such as BB1 (Wimperis 1994; Cummins \textit{et al.} 2003), which can suppress various sources of systematic error, are well known and fairly straightforward to implement (Xiao \& Jones 2006).

The situation is somewhat more complex in homonuclear systems, as two or more qubits of the same nuclear species will be affected by a single control field, making it much more difficult to perform single qubit gates selectively. While it is possible to use shaped soft pulses (Freeman 1998) to achieve sufficiently fine frequency excitation, these are inevitably much longer than the usual hard pulses, and it is no longer possible to neglect the spin--spin coupling terms during such pulses, introducing a significant source of error into the computation. It is also difficult to use composite pulses in this case, as these are even longer than ordinary soft pulses. A better approach is to use numerical techniques to simulate the evolution of the full system, and identify suitable control sequences from that.

\section{Controlling a Quantum System}

At the most basic level, the evolution of a quantum system can be described by the Schr\"{o}dinger equation
\begin{equation}
\Ket{\dot\psi} = -\ri \left( \mathcal{H}_0 + \sum_{k=1}^m u_k(t)\mathcal{H}_k \right)\Ket\psi
\end{equation}
where we are working in natural units, so that $\hbar=1$.  We divide the governing Hamiltonian into two distinct parts: $\mathcal{H}_0$ represents the internal couplings of the system which constantly act on it, and the set $\mathcal{H}_k$ are the possible external Hamiltonians which may be applied. The experimenter has the power to choose the functions $u_k(t)$ which describe the strength with which each of these control Hamiltonians acts as a function of time. The control problem is then to identify an appropriate set of $u_k(t)$, such that the resulting evolution of the system is as close as possible to the particular logic gate desired. This can be achieved by choosing a \it{fidelity} \rm function $\Phi(u_k)$, which is maximised when the correct logic gate is performed by a set of control functions, and then using standard optimisation techniques to locate this maximum and hence a sequence that performs the desired gate.

\subsection{Defining the Fidelity}

The first step towards this goal is to discretise the control fields $u_k(t)$ into control vectors $u_{kj}$, where we require that within a timestep $j$ the field strengths are all kept constant. Note that restricting ourselves to such a discrete form for the $u_k$, is not a serious issue, as any continuous control field function can always be characterised by a discrete approximation if enough timesteps are used. It is also the case that most experimental implementations of a gate will require such discretised control vectors, rather than the more general control functions initially described, so this approach has the added advantage of translating more easily into implementation.
This discretisation makes the Schr\"{o}dinger equation particularly straightforward to solve, giving
\begin{equation}
\Ket{\psi(t+\delta t)} = U\Ket{\psi(t)}
\end{equation}
where the propagator $U$ is given by
\begin{equation}
U = \re^{-\ri\delta t\left( \mathcal{H}_0 + \sum_{k=1}^m u_k\mathcal{H}_k \right) }
\end{equation}
It is then clear that we can simply concatenate these individual matrices to identify a final propagator $U_f$.
\begin{equation}
U_f = U_nU_{n-1}\dots U_2U_1
\end{equation}

Once a propagator has been calculated for a set of control vectors, all that remains is to choose a function to compare it with the target propagator $U_t$. A standard method for comparing two propagators is the Hilbert-Schmidt inner product
\begin{equation}
\Braket{U_t}{U_f} = \frac{\mathrm{tr}( U_t^\dagger U_f)}{N}
\end{equation}
with $N$ the dimension of the system. As we wish our metric to be independent of any global phase difference between the two propagators, we consider instead the square modulus of the inner product
\begin{equation}
\Phi = \Braket{U_t}{U_f}\Braket{U_f}{U_t}
\end{equation}
The function $\Phi$ defines a surface on the $n$ dimensional space described by the control vectors $u_k$, so that the better the logic gate, the higher the fidelity, up to a maximum of unity. We now consider how to search for these high quality sequences.

\subsection{Evaluating the Gradient}

The gradient is an extremely important component in many of the most efficient maxima finding approaches. The simplest way to obtain the gradient is to approximate it using finite difference methods. However, this requires at least $n+1$ evaluations of the fidelity function for each gradient calculation. As each fidelity evaluation requires the calculation of $n$ subpropagators this approach for obtaining the gradient scales with $n^2$ and very quickly grows impractical as $n$ increases.

An early attempt to address this issue led to the technique of ``strongly modulated composite pulses'' (Fortunato \textit{et al.} 2002). This technique keeps the number of time steps low (generally between 10 and 30), but allows variation in both step duration and frequency of the control fields in addition to the conventional control of amplitude and phase. Gradient calculation times remain reasonably short, but the system can have difficulty finding high fidelity pulses in more complicated systems. Experimentally, the sharp discontinuities between pulses can cause transient effects, both in pulse generation and in the probe circuit, thus reducing the practical value of the sequences (Ryan \textit{et al.} 2008).

A more sophisticated approach, GRadient Ascent Pulse Engineering or GRAPE, was subsequently introduced by Khaneja (Khaneja \textit{et al.} 2005; Schulte-Herbr\"uggen \textit{et al.} 2005). They observed that each evaluation of the fidelity only changes a single control pulse, and therefore only one subpropagator needs to be recalculated.
During the initial calculation of each subpropagator, we can also calculate and store both the forward and backward combined propagators as
\begin{align}
X_j &= U_jU_{j-1}\cdots U_2U_1\\
P_j &= U_{j+1}^\dagger\cdots U_N^\dagger U_t
\end{align}
Then when taking a derivative we separate out the relevant component
\begin{align}
\PD{\Phi}{u_{kj}} &= \PDO{u_{kj}}\left(\Braket{P_j}{U_jX_{j-1}} \Braket{U_jX_{j-1}}{P_j}\right)\nonumber\\
&= \Braket{P_j}{\PD{U_j}{u_{kj}}X_{j-1}}\Braket{X_j}{P_j} + \mathrm{c.c.}
\end{align}
This technique of storing the combined propagators during the fidelity evaluation stage makes it possible to evaluate the gradient with only the evaluation of $n$ further subpropagators. This is an example of the well known tradeoff between speed and memory use. For current systems and problems more than enough memory is available, making this an extremely effective technique.

This is a general approach which allows the gradient to be calculated efficiently irrespective of the precise form of each subpropagator. In the most common implementation where each timestep is short relative to the power of the Hamiltonians, we find that the first order approximation to the time derivative of $U_j$ is
\begin{equation} \PD{U_j}{u_{jk}} = -i\delta t\mathcal{H}_kU_j\end{equation}
giving the corresponding gradient element
\begin{align}
\PD{\Phi}{u_{kj}} &= -\Braket{P_j}{i\delta t\mathcal{H}_k X_j} - \mathrm{c.c.}\nonumber\\
&= -2\mathrm{Re}\Braket{P_j}{i\delta t\mathcal{H}_k X_j}\Braket{X_j}{P_j}
\end{align}

\subsection{Conjugate Gradients}

In non-linear optimisation problems where the gradient is known, one of the most powerful techniques is conjugated gradients (Shewchuk 1994). This method is guaranteed to find the maximum of a quadratic function in only $n$ steps, where $n$ is the dimension of the search space. It cannot in general achieve this convergence with a more complicated function such as the GRAPE fidelity, but in the vicinity of a maximum most functions can be approximated as quadratic using a Taylor series, and so the algorithm performs better the closer it gets to a maximum.

\subsection{Choosing an Initial Sequence}

Every search for an optimal control sequence has to start with an initial sequence which is then iteratively improved.  In order to encourage preferential development of smoothly varying control sequences, to facilitate experimental implementation, we use initial sequences composed of a superposition of sinusoids of different frequencies.

One major issue with the GRAPE approach is that in general it finds only local maxima, rather than global maxima, and so the search can become stuck on local maxima with inadequate fidelity. This is a particular problem when using composite search spaces, as described in the next section. In order to address this problem, the initial sequence is usually randomly generated; thus if one initial sequence fails to produce a suitable final result, another search can be performed using a different seed value.  However, rather than starting from a new seed value every time, it can be more efficient to perturb the optimised sequence a small distance in a random direction from the maximum and hope that it moves out of the region of that maximum and can move towards another, higher fidelity one. This technique can often lead to significant improvements in the fidelity with many fewer steps than would be required by resetting to a new seed, and having to optimise a whole new sequence.

\section{Extensions to the GRAPE technique}

The standard GRAPE algorithm is an effective and fast way to generate high fidelity pulse sequences in a wide range of systems. However, by changing the way we define the fidelity function, we can expand the types of sequence we can generate considerably.

\subsection{Composite Search Spaces}

Many experimental systems contain systematic errors which can interfere with the correct implementation of GRAPE pulse sequences. In NMR two of the most common (Cummins \textit{et al.} 2003) are Pulse Length Errors (PLEs) and Off Resonance Errors (OREs).  PLEs occur when the RF control field strength generated is systematically different to that requested, leading to a rotation angle which is either larger or smaller than expected. This behaviour is usually caused by inhomogeneity in the control field across the sample, so that a range of field strengths from about 70\% to 130\% of the nominal value is observed, and the final spectrum includes contributions from all these elements.  OREs appear when the frequency of the control field deviates from the expected value, changing the rotation axis. In NMR these errors can also arise due to variations across the sample, this time in the main field, giving rise to a range of Larmor frequencies.

Evaluating the fidelity of a pulse sequence in the presence of a known systematic error is easy, as it just requires modifying the Hamiltonian in the appropriate way. This also allows us to optimise a pulse sequence to take into account a particular strength of error. However, this is not enough. What we would like to do is to develop a sequence for which the effects introduced by the systematic error largely cancel out, producing essentially the same evolution as a system with no errors. More importantly, we need this to be the case for a whole range of systematic error values at the same time. This is made possible by using composite search spaces.

Rather than evaluating the fidelity once, with a particular strength of systematic error, we evaluate it multiple times, each time with a different strength of error, and sum these to produce an average fidelity. This produces a composite search landscape, where the remaining high fidelity points will perform well for a whole range of systematic errors. As the composition is purely linear, the gradient for the composite landscape can be found by summing the individual gradients.

The requirement to evaluate the fidelity multiple times per composite fidelity means the calculation time grows linearly with the number of error values being considered. However, the composite nature of the search space can also greatly increase the number of local maxima of insufficiently high fidelity. It is important therefore to find a balance between the range of errors to be tolerated, the speed of calculation and the final quality of result.

Composite search spaces will identify sequences which perform well with a set of precise systematic errors. In most cases this will correspond to comparably high performance for the range of values in between, but in some cases the dynamics of the system allow the generation of sequences which perform extremely well at the specified error values, and extremely poorly in between. The risk of such behaviour can be reduced by avoiding choosing periodic intervals between error values.

\subsection{Suppression of Contaminant Spins}
It is common in NMR that a sample contains a small quantity of additional impurity molecules containing spin-$\half$ nuclei of the same nuclear type as the computational spins. In particular, samples dissolved in D$_2$O will inevitably contain a greater or lesser quantity of only partially deuterated HOD. Furthermore, unless a sample is extremely well sealed, this situation will only deteriorate over time (Derome 1987).

Because the extra qubit is on a different molecule to the computational qubits, there is no direct coupling between them, and there will be no interference at the computation stage. However, the readout signal will contain additional peaks, sometimes with a magnitude significantly in excess of the primary spins, and in an arbitrary state. A fully dispersive water line can significantly distort the spectrum more than a kHz away from its centre frequency.  A wide range of techniques have been developed for suppressing such signals (Hore 1983), but these are not always suitable for use in quantum computing experiments.

A variation of the composite search spaces technique can be applied to solve this problem. The contaminating system will generally be only a single isolated spin, and the effect of the control radiation upon it can be simulated in precisely the same way as for the primary qubits. A fidelity between this evolution and the identity gate may then be calculated, and this impurity fidelity combined with the main fidelity as described above. This will lead the algorithm to generate sequences which will not only perform the correct logic gate on the computational system, but will not excite the impurity spin, preventing it from interfering at the readout stage.

\subsection{Pulses and Delays}
Some quantum logic gates, such as the controlled-\textsc{not}, rely on the internal Hamiltonian to drive the desired evolution. As this is often a small coupling, such gates are generally relatively long, sometimes up to two orders of magnitude longer than simple single qubit gates. Two basic approaches to deal with this are to increase the number of time steps in a sequence, or to increase the length of each time step. Increasing the number of time steps will proportionally increase the time required for each fidelity and gradient evaluation, which soon becomes impractical. Using longer time steps keeps the calculation time short, but does prevents the sequence from exhibiting high frequency behaviour, which is often important in producing high fidelity sequences. Furthermore, as the time steps grow longer, the approximations used to calculate the derivative will become inaccurate.

Traditionally, these sequences have often been implemented using short pulses combined with delays to allow the coupling to affect the system, and basic GRAPE techniques can be used to generate these short pulses.  There will, however, be errors in each pulse, which will accumulate through the sequence.  Additional errors can also occur in experimental implementations, arising from additional delays between pulses, as discussed in \S3$\,b$ below. Instead it is better to calculate the entire gate in one step, this canceling, as far as possible, the errors from each component.

So far we have only considered one form of subpropagator, a fixed length pulse with a set of control Hamiltonians applied. However, there is no need to restrict ourselves to this form. A useful second subpropagator to is a variable length delay, with no control Hamiltonians applied. It is easy to calculate the form of this subpropagator and its gradient
\begin{align}
U(t) &= e^{-i t H_0}\\
\OD{U(t)}{t} &= -i\, H_0 \,U(t)
\end{align}
and it is straightforward to include subpropagators of this form directly in a GRAPE sequence alongside the standard ones, where the parameter $t$ takes the place of the $u_k$. This demonstrates a more general technique, as any subpropagator for which $U$ and $\PD{U}{u_i}$ can be expressed in matrix form can be included in the GRAPE algorithm while maintaining the ability to calculate the total gradient in linear time.

\section{Experimental Implementation Details}
\label{sec:experimentaldetails}
Using the techniques described above, it is possible to generate high fidelity sequences to perform a wide range of different logic gates, including spin selective gates in homonuclear systems, controlled-\textsc{not} type gates and precise null (\textsc{identity}) gates. However, it must always be remembered that these gates are being designed for use in an experimental context, and the final judgement on their quality must be reserved until they are tested on the target equipment.

There are many factors which can interfere with the precise implementation of the simulated sequence, and some of these may produce a significant deterioration in the performance of the experimental sequence. We review here some of the more significant factors we have been forced to consider.

\subsection{Rounding Errors}
Any experimental implementation will necessarily have some limits on the precision with which control fields can be applied. In some cases these will be absolute limits, in others there is a relative precision where only a fixed number of points are possible between 0 and some maximum value. In both cases there is a risk that the inevitable rounding that must be performed will cause a significant reduction in the quality of the sequence. This effect is easy and quick to simulate, and so it is generally straightforward to confirm whether a sequence is likely to be seriously affected. 

\subsection{Additional Delays}
When performing a control sequence, the hardware may well introduce additional delays around the actual sequence, for example to allow amplifier gating, or to conform to a set timing resolution. The evolution of the system during these periods will change the effect of the sequence. In many cases this will be only a very small error, but in complicated programs involving many control sequences, these errors can accumulate and become significant.

These additional delays can easily be included in the simulation of the GRAPE pulse; in practice, however, the most straightforward way to remove this source of error is to adjust the target propagator for the sequence instead, so that the combined action of the target propagator and the surrounding delays produces the desired result.

\subsection{Transients}
Whenever the strength of a control Hamiltonian is changed during a control sequence, there will inevitably be some kind of transient generated as the system transitions from one strength to another. Our GRAPE sequences are encouraged to form smoothly varying shapes to reduce the shifts between steps but, due to the way control fields are generated, significant transients can sometimes still occur. It then becomes important to ensure that the length of each time step is somewhat longer than the die down time of the transients. An effective way to identify the shortest practical time step is to develop a set of sequences, all of the same total duration, but with different time steps.  This is easily achieved by forming each new sequence by combining pairs of time steps in the previous sequence into a single period and then re-optimising the new sequence.

Figure \ref{fig:transientfigure} shows an example of this behaviour: each increase in the number of time steps produces a theoretical improvement, as the shorter time steps theoretically allow better control, but there is a sharp decrease in experimental performance once the time step length drops below a threshold value.  To avoid these effects sequences should normally be designed using times steps at or above this threshold value.

\subsection{Strong Coupling}
In traditional NMR quantum computing it is convenient to assume that the coupling between spins is of Ising ($zz$) form, which is reasonable when the coupling strength $J$ is much smaller than the separation of the spin frequencies. This is usually called the weak coupling approximation, and as the name implies is only approximately correct in homonuclear spin systems (in heteronuclear systems the approximation is essentially perfect). However, one of the advantages of GRAPE is that it is capable of handling any form of $\mathcal{H}_0$, so gates can be developed which use the exact form of the underlying coupling interaction, taking any additional evolution into account and preventing small errors building up over multiple gates. 

\subsection{Maximum Power}
In any experimental setup there will be physical limits to the strength with which the desired control Hamiltonians can actually be applied. It is important to make sure that GRAPE, or any other technique for designing sequences, does not generate sequences which cannot actually be implemented. A power limit can be enforced by the addition of a penalty function which acts to reduce the ``fidelity'' if the power requirement goes too high.

If the penalty function is an additive one then the gradient is simple to calculate, being just the sum of the two gradient functions. However, penalty functions are not entirely simple to implement and they introduce a range of complications. The penalty function has to be severe enough to provide a hard discouragement above the maximum power, but should not have too great an effect at acceptable powers to avoid driving the algorithm away from potential candidate sequences near the limit. This means a steeply varying penalty which may lead to an extremely narrow peak in a few search dimensions. The result is that the algorithm may be unable to take large steps along the gradient and the search can become bogged down in low fidelity areas.

Before attempting to tackle these problems, it is worth experimenting with generating sequences to see whether the power is in fact going above maximum. In our experience, many logic gates have their power range limited by the size of the internal couplings, and a careful choice of low power seed sequences can also have a restricting effect on the growth of sequence power.  The use of a composite search space can also be useful, as the effects of PLEs can become very substantial at high power levels.

\section{Experimental Implementation Results}
\begin{figure}[t]
\centering
\includegraphics[width=6cm]{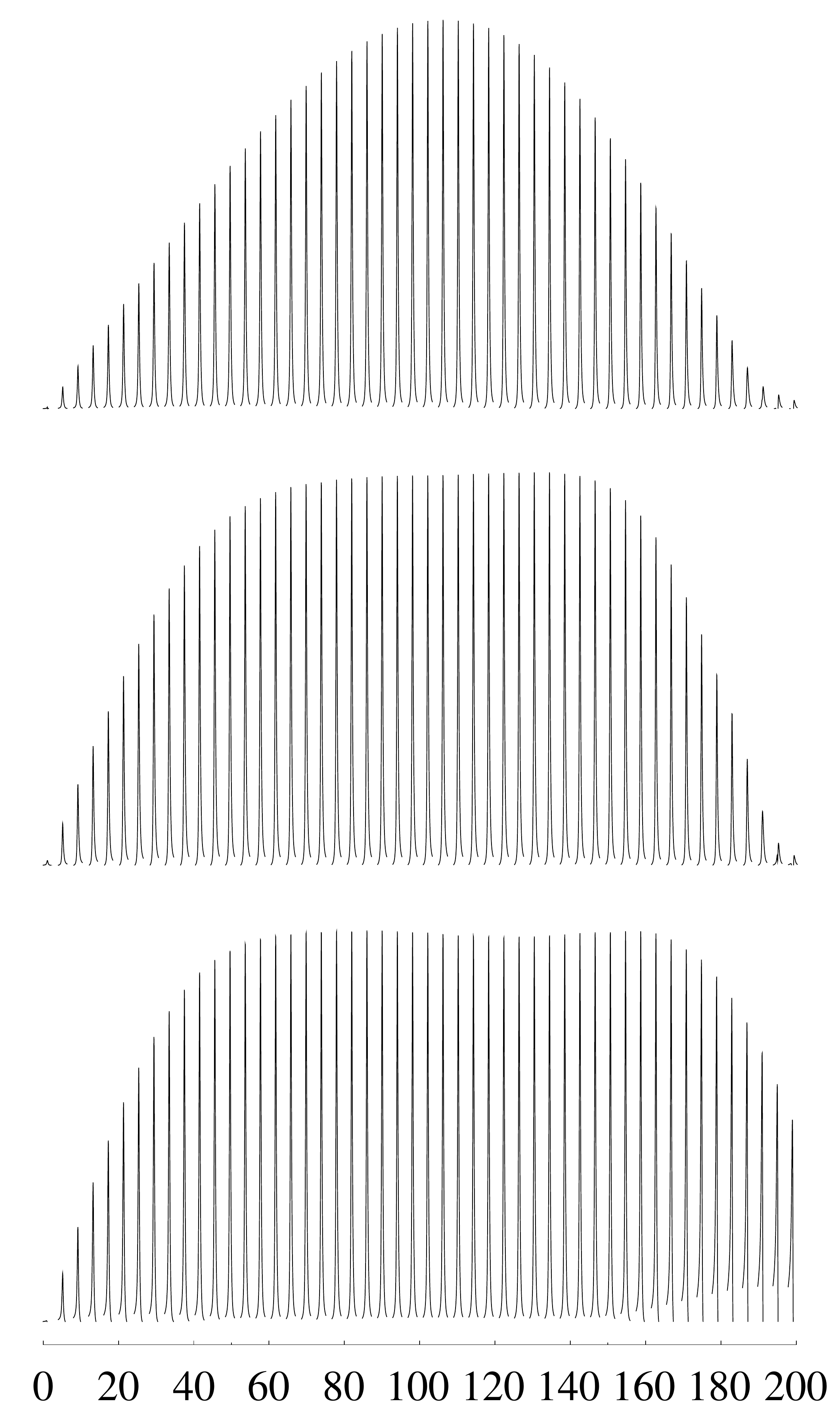}
\rule{35em}{0.5pt}
\caption[Error Tolerant Gates]{Each figure shows 50 individual spectra from a system containing a single isolated spin. Each spectrum was acquired with a different artificially induced pulse strength in the range 0-200\% of nominal, corresponding to errors in the range $\pm100\%$. The top figure shows the sine wave variation in the NMR response produced by a simple hard pulse. The middle figure uses BB1, which removes error terms up to the sixth power. The bottom figure is from a GRAPE sequence, and exhibits a fairly flat response over a range of strengths even wider than for the BB1 version.}
\label{fig:BB1beater}
\end{figure}

In this section we present a few results from our experiments with the GRAPE algorithm. As mentioned above, GRAPE is particularly important in homonuclear systems, where conventional hard pulses and composite pulse techniques cannot easily be used. However, using composite search spaces, GRAPE can also be used to develop error tolerant pulses which may be useful in some heteronuclear systems as well. Standard composite pulses such as BB1 (Wimperis 1994; Cummins \textit{et al.} 2003; Xiao \& Jones 2006) already provide extremely high tolerance of the most common errors in NMR, but it is possible that GRAPE could be more important in systems where other types of error are also important. As a demonstration of this application of GRAPE, we sought to develop single qubit sequences which could compete with BB1 for resistance to PLEs. Figure \ref{fig:BB1beater} shows an example of this for a 90$^\circ$ rotation gate, where a GRAPE sequence with a even wider tolerance of pulse length errors than BB1 is demonstrated.

Of course, our primary motivation to use GRAPE is to gain access to homonuclear or mixed systems, and so we now present some further results from experiments with homonuclear two spin systems. There are techniques available to apply heteronuclear techniques to a two spin homonuclear system, such as applying stroboscopic pulses on resonance with the spin--spin coupling term (Jones \& Mosca 1999), but this is impossible to scale into larger systems.  An alternative is to use a compiler approach to keep track of the effects of the interactions (Bowdrey \textit{et al.} 2005), but this approach does not normally permit two or more logic gates to be performed in parallel.  GRAPE suffers from none of these drawbacks, and any desired unitary transform can be generated.  It is also relatively straightforward to apply to a system of arbitrary size, although the computational complexity scales badly with the number of qubits involved (Ryan \textit{et al.} 2008). It is, therefore, useful to use a small system as an initial testbed, as the calculations are quicker and less complex.

Our two spin GRAPE experiments use a 50\,mM solution of the pyrimidine base cytosine in D$_2$O at a temperature of $25^\circ$C; a rapid exchange of the two amine protons and the single amide proton with the deuterated solvent leaves two remaining protons forming an isolated two spin system (Jones \& Mosca 1998), accompanied by a fairly large HOD impurity signal.  Following NMR conventions the two coupled spins, which are the qubits of interest, are referred to as $I$ and $S$.  All experiments were performed on a Varian Unity INOVA NMR spectrometer with a nominal $^{1}\textrm{H}$ frequency of 600\,MHz, and the spectrometer frequency placed on resonance with the residual HOD signal.

One of the most important abilities of the GRAPE algorithm is the generation of arbitrary unitary propagators. Conventional shaped pulses (Freeman 1998) often produce ``point-to-point'' sequences, which will transfer a system from one known state to another.  In contrast a GRAPE derived sequence will perform a genuine logic gate, which will transform all states correctly, independent of the starting state of the system. This can be confirmed by performing full quantum state tomography to fully characterize the transformation matrix, but this is an arduous process; for brevity we have adopted a more compact test of sequences, simply observing their action on spin states aligned with the major axes of the system. This demonstrates the flexibility of the GRAPE sequence without the full rigour of state tomography.

\begin{figure}[htbp]
\centering
\includegraphics{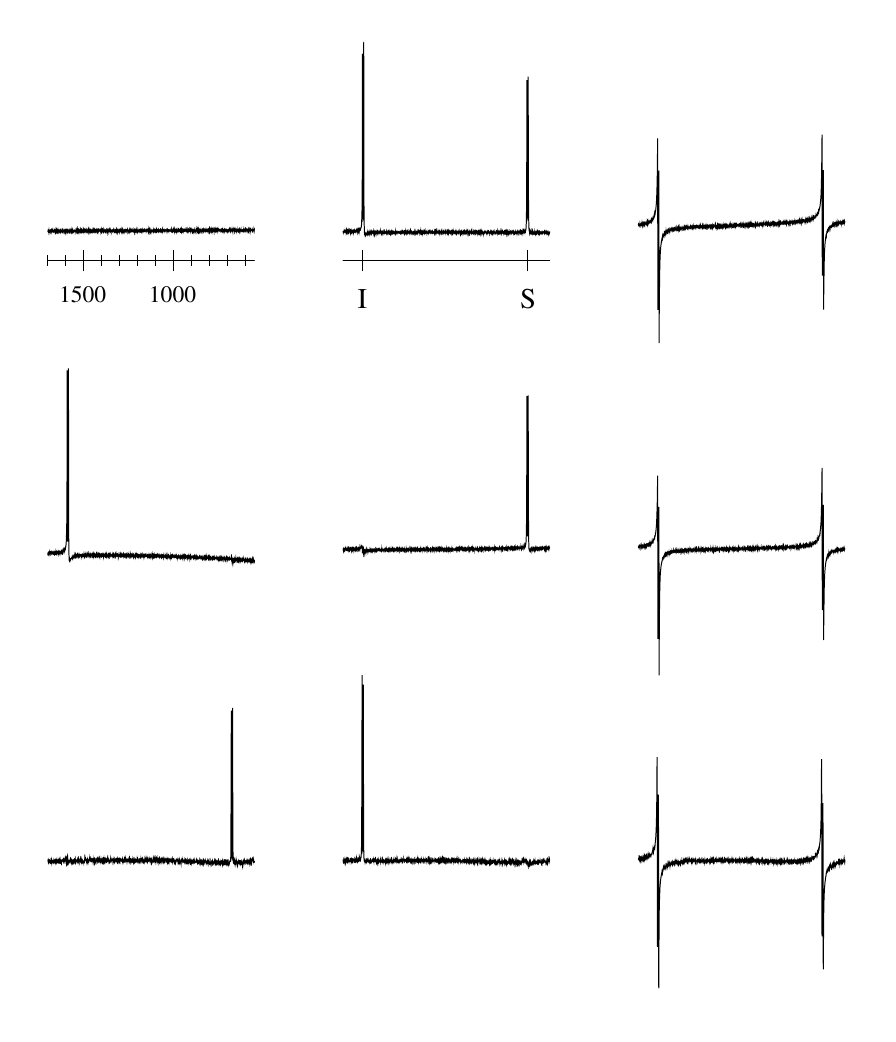}
\rule{35em}{0.5pt}
\caption[Transient Issues]{GRAPE derived sequences are true logic gates which are independent of starting state and are not simple ``point-to-point'' transformations. The top row of spectra show signals from states prepared with both qubits aligned along the $z$, $x$ and $y$ axes of the Bloch sphere respectively. There is no signal from the $z$ state as there is no transverse magnetisation, and only noise is observed. The $x$ and $y$ signals are very similar, being simply 90$^\circ$ out of phase with each other. The axis below the $z$ state spectrum shows the frequency scale in Hz; frequencies are measured from the reference frequency of the NMR spectrometer, which was chosen to place the residual HOD signal (not shown) on resonance at zero frequency.  Following NMR conventions this axis is plotted with frequencies increasing from right to left, and the intensity axis, which is in arbitrary units, is not shown.  Below the spectrum from the $x$-state we indicate schematically which signals arise from the two spins $I$ and $S$ respectively. The second and third rows then show the resulting state after a GRAPE derived selective 90$^\circ$ rotation gate about the $y$-axis is applied to the I qubit (middle row) and S qubit (bottom row). As expected in all cases there is no change for the un-targeted spin. From an initial $z$ state, the targeted spin is excited into the $x$ state, while starting from the $x$ state, the spin ends up in a $-z$ state and no signal is visible. Rotation about the $y$-axis leaves a $y$ state unchanged, and so all three $y$ spectra are essentially unchanged. }
\label{fig:trueI90andS90}
\end{figure}

\begin{figure}[htbp]
\centering
\includegraphics{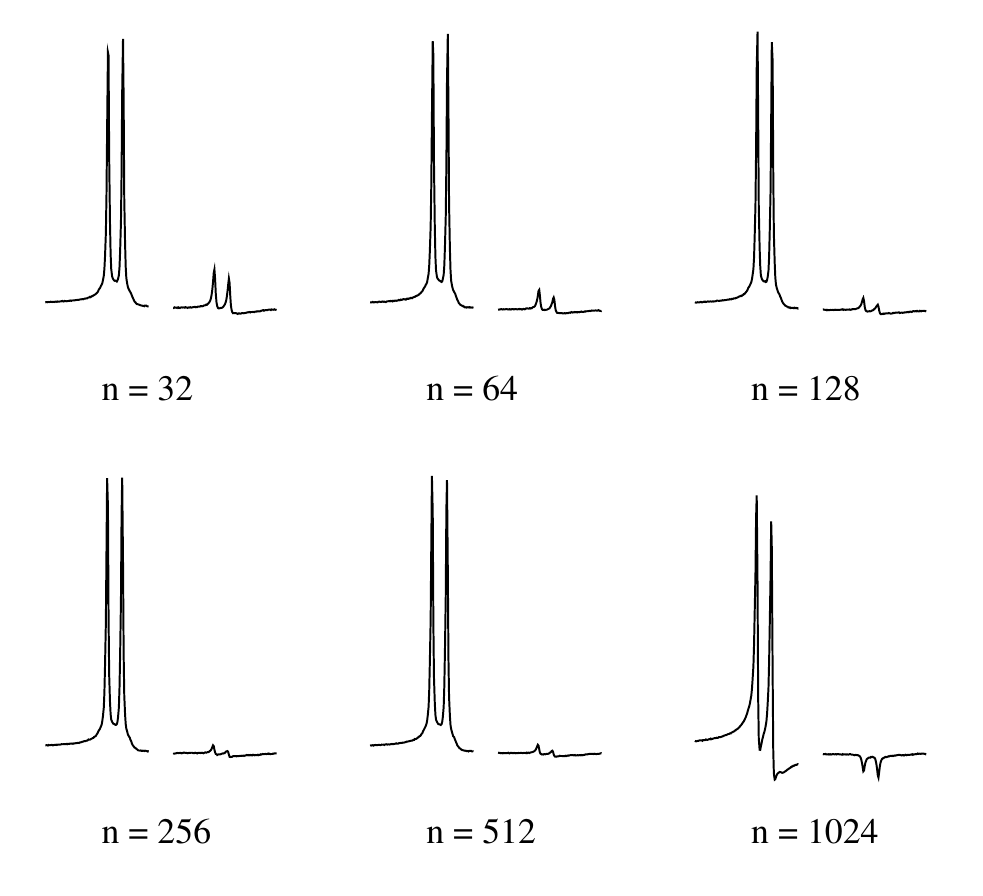}
\rule{35em}{0.5pt}
\caption[Transient Issues]{A high fidelity sequence performing a selective 90$^\circ$ rotation on the I (left) qubit, lasting 2048\,$\mu${s} and comprising 1024 time steps was used to prepare a set of progressively coarser pulses. Each sequence was prepared by averaging pairs of time steps from the previous sequence, to produce a sequence with half the number of time steps, and then re-optimising with GRAPE to the highest achievable fidelity. For compactness here only the components of each spectrum close to the peaks are shown. As the number of time steps increases, the quality of the sequence clearly improves, with less and less excitation on the un-targeted spin. However, at the transition between $n=512$ and $n=1024$, there is a sudden and substantial deterioration in the quality of the final experimental result, even though the more complex sequence has a higher theoretical fidelity. This result has been confirmed for a variety of other sequences (data not shown) and seems to indicate a clear threshold beyond which transient effects between time steps give rise to an unacceptable error. This results in a hard limit on the duration of time steps in generating GRAPE sequences, although the exact limit will be a property of a particular experimental setup, and will need to be determined individually for each system.}
\label{fig:transientfigure}
\end{figure}

The results of such an experiment for two different selective rotation gates are shown in Figure \ref{fig:trueI90andS90}.  These results seem fairly promising, demonstrating high quality selective gates derived using GRAPE. However, for practical quantum computations, very high gate fidelities are required (Ryan \textit{et al.} 2008), and it is impossible to analyse the effects of a sequence to sufficient accuracy using these spectra.  Figure \ref{fig:transientfigure}, which shows the effects of transients, uses a cut down form of spectrum to allow much greater levels of detail to be seen, by only displaying 25\,Hz either side of each qubit. In this figure, it is easy to see that there is some excitation of the un-targeted spin in a selective gate, and we have so far been unable to remove this entirely. It also appears to be very difficult to control the rotation angle and phase of the selective rotation to better than a few degrees.

\begin{figure}[htb]
\centering
\includegraphics{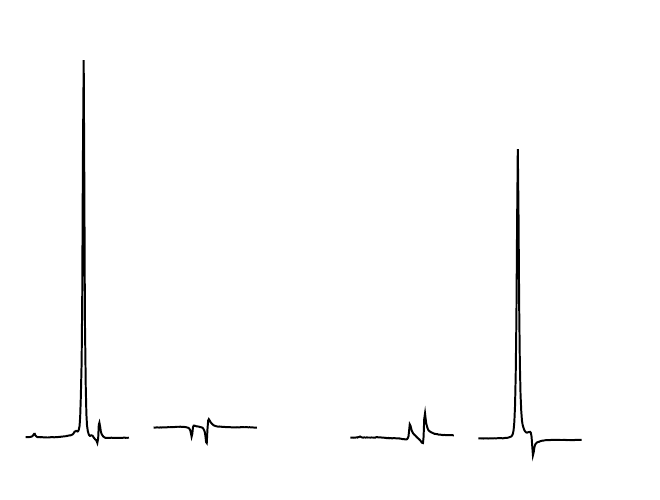}
\rule{35em}{0.5pt}
\caption[Pseudo-pure state preparation]{A pseudo-pure state prepared using controlled-transfer gates is observed by selective excitation of either the I spin (left figure) or S spin (right figure). As for Figure \ref{fig:transientfigure}, a cut down spectrum only showing the frequencies around each spin is used to show greater detail within each multiplet. In each case, only the left line of the observed doublet is visible, showing that the unobserved spin is almost entirely in the $\Ket{0}$ state as expected. There is a small amount of residual signal from the $\Ket{1}$ state, and some excitation of the un-targeted spin, which may be attributed to cumulative errors through the multiple gates in the sequence.}
\label{fig:PPS}
\end{figure}

The most effective way to test the quality of these sequences more accurately is to use them in more complicated algorithms involving multiple gates, and then observe any resulting errors. We chose to do this using the controlled-transfer gate method of pseudo-pure state preparation (Kawamura \textit{et al.} 2010). This is a good algorithm to use because the ability to prepare pseudo-pure states accurately is itself an important part of any quantum computation. It can be implemented using only two types of selective logic gate, with five actual gates being used, and is known to work well in heteronuclear spin systems. The results of these experiments are shown in Figure \ref{fig:PPS}.

\section{Remaining Errors}
 These results show that there is still some way to go before our implementation of GRAPE sequences can supply the extremely high fidelity propagators required for complex quantum computation algorithms. The simulator predictions for the sequences shown in Figure \ref{fig:PPS} show essentially perfect creation of the pseudo-pure state, but this is not born out in practice.
 
 Although several of the most significant causes of experimental error have been addressed, as described in Section \ref{sec:experimentaldetails}, there remain a number of smaller issues which may be causing the residual errors in the experimental results. These include transient effects, small areas of the sample with extreme RF inhomogeneity and small deviations in the relative angle of different pulses generated by our spectrometer. These kind of flaws are all highly implementation dependent, and so it is important for researchers using GRAPE style pulses to characterise their own equipment as completely as possible to identify potential sources of error in the system.  A further potential source of errors is decoherence during pulses (Wu \textit{et al.} 2011), and especially during long delays within sequences of pulses and delays, but for the simple pulses considered above this is unlikely to be a major problem.

 There is, however, an additional source of experimental error, which may be dominant in our system.  So far we have assumed that the spectrometer can produce any desired amplitude and phase of RF radiation with equal accuracy.  Whether this is in fact likely to be the case will depend on exactly how the spectrometer synthesises shaped pulses.  Many modern spectrometers use direct digital synthesis (Yun \textit{et al.} 2002), but older systems, including ours, use analogue techniques, and are prone to errors arising from this.  This can be investigated by mixing the shaped pulse RF with a pure frequency source, to beat the RF down to zero frequency, and then observing the result directly with a digital oscilloscope.

 Initial studies of our system (data not shown) suggest that control of RF amplitude at a fixed pulse phase is highly linear, and is unlikely to be a source of error, but the use of small angle phase shifts leads to small but significant errors in both the amplitude and phase of the generated RF.  These can arise as small angle phase shifts are generated by combining appropriate amounts of two separate frequency sources generated in quadrature.  If these two underlying sources are either not exactly in quadrature, or alternatively if their amplitudes are not exactly the same, then the result will be systematic errors in RF amplitude and phase, with a size depending on the phase of the RF being generated.

 Our initial measurements are consistent with an error of this kind, which may be sufficient to explain the relatively poor performance of GRAPE in our experimental system.  If this is correct, then in principle errors of this kind could be calibrated and corrected for within the implementation.  A full resolution of this question will, however, require further work.

 \section{Conclusions}

 The GRAPE algorithm provides the ability to derive shaped pulse sequences to perform arbitrary logic gates, even in homonuclear or hybrid systems, far more simply than was previously possible using standard techniques. By making various additions to the core GRAPE concepts, it is also possible to obtain sequences which are comparatively robust against certain types of experimental error.

 For simple results examining only the performance of individual logic gates, it appears that generally high performance can be achieved, provided the equipment capability falls within the tolerance window of the sequence. However in more complicated algorithms where more than one logic gate is applied, cumulative errors from the multiple gates mean that in our implementation the quality of the logic gates is not yet at the standard required for complex quantum algorithms. Our theoretical results show substantial promise; if experimental techniques can match the required standards for implementing sequences, GRAPE will certainly have a useful part to play in demonstrating new algorithms and quantum phenomena.

\end{document}